\documentclass[aps,pra,twocolumn,groupedaddress,showpacs,showkeys]{revtex4}

\usepackage{graphicx}
\begin{document}
\vspace{1cm}
\newcommand{\be}{\begin{equation}}
\newcommand{\ee}{\end{equation}}

\title{Variations of Casimir energy from a superconducting transition}

\author{ Giuseppe Bimonte, Enrico Calloni,  Giampiero Esposito and Luigi Rosa}
\affiliation{Dipartimento di Scienze Fisiche, Universit\`{a} di
Napoli Federico II,   Via Cintia I-80126 Napoli, Italy; INFN,
Sezione di Napoli, Napoli, ITALY }

\date{\today}

\begin{abstract}
We consider a five-layer Casimir cavity, including a thin
superconducting film. We show that when the cavity is cooled below
the critical temperature for the onset of superconductivity, the
sharp variation (in the microwave region) of the reflection
coefficient of the film  produces a variation in the value of the
Casimir energy. Even though the relative variation in the Casimir
energy is very small, its  magnitude can be comparable to the
condensation energy of the superconducting film, and thus causes a
significant increase in the value of the critical magnetic field,
required to destroy the superconductivity of the film. The
proposed scheme might also help clarifying the current controversy
about the magnitude of the contribution to Casimir free energy
from the TE zero mode, as we find that alternative treatments of
this mode strongly affect the  shift of critical field.

\end{abstract}


\maketitle

\section{Introduction}

In recent years, new and exciting advances in experimental
techniques \cite{decca} prompted a great revival of interest in
the Casimir effect, over fifty years after its theoretical
discovery \cite{casimir} (for a recent review on both theoretical
and experimental aspects of the Casimir effect, see Refs.
\cite{bordag}-\cite{lamor}). As   is well known, this phenomenon
is a manifestation of the zero-point fluctuations of the
electromagnetic field: it is a purely quantum effect and it
constitutes one of the rare instances of quantum phenomena on a
macroscopic scale.

In his famous paper, Casimir evaluated the force between two
parallel, electrically neutral,  perfectly reflecting plane
mirrors, placed a distance $L$ apart, and found it to be
attractive and of a magnitude equal to: \be F_{C}=\frac{\hbar c
\pi^2 A}{240 \;L^4}\;.\label{paral}\ee Here, $A$ is the area of
the mirrors, which is supposed to be much larger than $L^2$, so
that  edge effects become negligible. The associated energy $E_C$
\be E_C=-\frac{\hbar c \pi^2 A}{720 \;L^3}\;.\label{epara}\ee can
be interpreted as representing the shift in the zero-point energy
of the electromagnetic field, between the mirrors, when they are
adiabatically moved towards each other starting from an infinite
distance. The Casimir force is indeed the dominant interaction
between neutral bodies at the micrometer or submicrometer scales,
and by modern experimental techniques it has now been measured
with an accuracy of a few percent (see Refs. \cite{decca} and
Refs. therein).

The issue of the energy of vacuum fluctuations, a manifestation of
which is provided by the Casimir effect, is   of the outmost
importance in other areas of Physics, most notably in Cosmology,
where it is known to raise a number of problems, when its
gravitational effects are considered \cite{weinb}. For this
reason, it is clearly very important to devise laboratory tests to
study the physical consequences of vacuum fluctuations, as a way
to check our theoretical understanding of this intriguing quantum
phenomenon.

All experiments on the Casimir effect performed so far measured
the Casimir force, in a number of different geometric
configurations. In  a recent Letter \cite{bimonte} we have found
that by realizing a rigid cavity,  including a superconducting
film, it should be possible to measure directly the variation of
Casimir energy accompanying the transition to the normal state of
the superconducting film, in an external magnetic field.

Apart from the fundamental interest of a direct measurement of a
(variation of) Casimir energy, rather than a force, our scheme has
the further advantage that, being based on rigid cavities, it
should  make it easier to study  the dependence of the Casimir
effect on the geometry of the cavity. Indeed,  since this effect
arises from  long-range correlations between the dipole moments of
the atoms forming the walls of the cavity, that are induced by
coupling with the fluctuating electromagnetic field, the Casimir
energy depends in general on the geometric features of the cavity.
For example, we see from Eq. (\ref{epara}) that, in the simple
case of two parallel slabs, the Casimir energy $E_C$ is negative
and is not proportional to the volume of the cavity, as would be
the case for an extensive quantity, but actually depends
separately on the area and distance of the slabs. Indeed, the
dependence of $E_C$ on the geometry of the cavity can reach   the
point where it turns from negative to positive, leading to
repulsive forces on the walls. For example (see Ref.
\cite{bordag}), in the case of a cavity with the shape of a
parallelepiped, the sign  of $E_C$ depends on the ratios among the
sides, while in the case of a sphere it has long been thought to
be positive. It is difficult to give a simple intuitive
explanation of these shape effects, as they hinge on  a delicate
process of renormalization, in which the finite final value of the
Casimir energy is typically expressed as a difference among
infinite positive  quantities. In fact, there exists a debate, in
the current literature, whether some of these results are true or
false, being artifacts resulting from an oversimplification in the
treatment of the walls \cite{barton}. Under such circumstances, we
think it would be desirable to have the possibility of an
experimental check of these statements, for the maximum possible
variety of configurations.

Another reason of interest in our superconducting cavities, not
discussed in  \cite{bimonte},  is that their study will help
clarifying one of the most debated issues in the current
literature on the Casimir effect, i.e. the question of the
contribution from the TE zero mode to thermal corrections in real
metals, i.e. characterized by a finite conductivity. At the
present time, there is no general agreement between the experts on
how to deal with this problem, and we address the reader to Refs.
\cite{geyer} for a survey of the extended literature on this
topic. A superconducting cavity is a very good tool to explore
this problem, because the variation of Casimir energy across the
transition precisely arises from the fact that the film has a
finite conductivity, that changes when it becomes superconducting.
Indeed, the computations presented in Sec. IV show that the
critical field required to destroy the superconductivity of our
cavity is very sensitive to the contribution of the TE zero mode,
at the level that the alternative treatments proposed in the
literature may lead to predictions that can differ even by 100
percent.

The plan of the paper is as follows. In Sect. II we present the
general scheme of superconducting Casimir cavities,   in Sec. III
we  discuss the applicability of Lifshitz theory \cite{lifs} to
the computation of the variation of Casimir energy in the
superconducting transition. In Sec. IV we  present the results of
our numerical computations, including in Subsec. C a study of the
contribution from the TE zero mode and finally, in Sec. V, we draw
some conclusions and outline possible future work. We have
included two Appendices to review the basic facts on type I
superconductors that are useful to understand the  properties of
our superconducting cavities.

\section{The superconducting cavity}

The Casimir cavity that we consider  is the  planar five-layer
system depicted in Fig. 1: a thin superconducting film of
thickness $D$ is placed between two thick metallic slabs (made of
a non-magnetic and non superconducting metal), that constitute the
plates of the cavity. The gaps of width $L$ separating the film
from the plates are supposed to be filled with some insulating
material.

We consider cooling the cavity at a temperature $T$ below the
critical temperature $T_c$ for the onset of superconductivity in
the thin film. A magnetic field parallel to the film is then
turned on, and its intensity is gradually increased, until it
reaches the critical value $H_{c\|}$, for which superconductivity
of the film is destroyed. The question we ask is   whether the
Casimir energy stored in the cavity  affects in an appreciable way
the value of $H_{c\|}$. To address this question (see Appendix I
for a short review of the magnetic properties of type I
superconductors), we start from Eq. (\ref{hfilm})  that connects
$H_{c\|}$ to the (density of) condensation energy $e(T)$ of the
film. Upon multiplying both sides of Eq. (\ref{hfilm}) by the
volume $V$ occupied by the film, we obtain: \be \,\frac{V}{8
\pi}\,\left(\frac{H_{c \|}(T)}{\rho}\right)^2 = {\cal E}_{\rm
cond}(T)\;,\label{hcri}\ee where ${\cal E}_{\rm cond}(T)=V\, e(T)$
is the condensation energy of the film.
\begin{figure}
\includegraphics{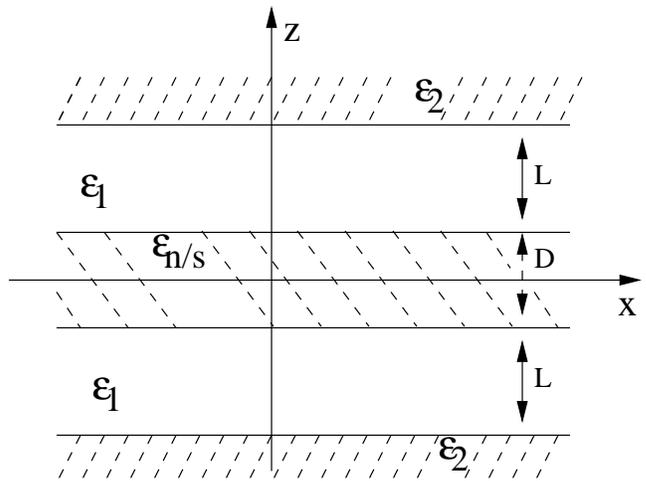}
\caption{\label{fig1} Scheme of the superconducting five-layer
cavity.}
\end{figure}
 When the film is included in a
cavity, the r.h.s. of the above Equation should be augmented by an
extra term, i.e. the $\Delta F^{(C)}_E(T)$ representing the {\it
difference} \be \Delta F^{(C)}_E(T):= F^{(C)}_n(T)-F^{(C)}_s(T)
\ee among the Casimir free energies of the cavity, in the
superconducting ($s$) and in the normal ($n$) states of the film
(in zero magnetic field). While we postpone to the next Section
the computation of  $\Delta F^{(C)}_E(T)$, we observe here that
the  presence of such a term  should be expected on physical
ground, because the transition to superconductivity causes a sharp
variation in the reflective power of a film in the microwave
region of the spectrum \cite{glover}, and this obviously produces
a change in the Casimir energy, which depends on the reflection
coefficients of the plates. Thus, for the cavity case, Eq.
(\ref{hcri}) should be replaced by \be \frac{V}{8
\pi}\,\left(\frac{H_{c \|}^{\rm cav}(T)}{\rho}\right)^2 \,= {\cal
E}_{\rm cond}(T)\,+\,\Delta F^{(C)}_E(T)\;.\label{hcricav}\ee In
writing this equation,  the assumption has been made that the
value of the coefficient $\rho$ does not change when a film is
placed in a cavity. This approximation amounts to saying  that all
quantities referring to the film, like the penetration depth,
condensation energy etc. are not affected by virtual photons in
the surrounding cavity. This is a very good approximation, since
the leading effect of radiative corrections is a small
renormalization of the electron mass \cite{kreuzer} of order
$\varepsilon := \alpha \times \hbar \omega_c/(m c^2)$ (up to
logarithmic corrections), where $\alpha$ is the fine structure
constant, $\omega_c=c/L$ is the typical angular frequency of
virtual photons, and $m$ is the electron mass. Even for $L$ as
small as 10 nm, $\varepsilon \approx 3 \times 10^{-7}$. The
associated shift of critical field is of the same order of
magnitude, and thus negligible with respect to that caused by
$\Delta F^{(C)}_E$, which will turn out to be of some percent (see
Sec. IV).

Equation (\ref{hcricav}) deserves  further comments. Consider
first the relative shift of parallel critical field: \be
\frac{\delta H_{c \|}}{H_{c \|}} := \frac{H_{c \|}^{\rm cav}-H_{c
\|}}{H_{c \|}}\;,\ee which, according to Eqs. (\ref{hcri}) and
(\ref{hcricav}), is equal to \be \frac{\delta H_{c \|}}{H_{c \|}}
=\sqrt{ \; \frac{\Delta F^{(C)}_E}{{\cal E}_{\rm cond}
}+1}-1\;.\label{shift}\ee  Now, the relative variation $\Delta
F^{(C)}_E/F_n^{(C)}$ of Casimir energy  should be expected a very
small number, because the Casimir energy is mostly associated with
photon energies  of order $\hbar \omega_c$, i.e.  $10 \div 20$ eV
for $L \simeq$ 10 nm, while the transition to superconductivity
affects the reflective power only at photon energies of order $k
\,T_c \approx 10^{-4}$ eV for $T_c \simeq 1$ K. However, and this
is the key point, Eq. (\ref{shift}) shows that {\it the shift of
critical field is determined by the variation of Casimir free
energy $\Delta F^{(C)}_E$ as compared to the condensation energy
${\cal E}_{\rm cond}$ of the film}.  The latter is  very small:
for a thin film  with, say, an area $A$ of 1 ${\rm cm}^2$ (the
area $A$ is not really important, because for $A \gg L^2$ both
${\cal E}_{\rm cond}$ and $\Delta F^{(C)}_E$ are proportional to
$A$) and a thickness of a few nm it is easily of  order $10^{-8}$
erg.  By contrast, we see from Eq. (\ref{epara}) that a typical
Casimir energy for a cavity with an area of 1 ${\rm cm}^2$ and a
width $L=10$ nm, has a magnitude of 0.43 erg, i.e. over seven
orders of magnitude larger than the condensation energy of the
film. This implies that even a relative variation of Casimir
energy as small as one part in $10^8$  would still correspond to
more than $10 \%$ of the condensation energy of the film, and
would induce a shift of critical field of over 5 \%.

The huge increase in sensitivity  that we   predict by virtue of
a second energy scale, the condensation energy, which is orders of
magnitude smaller than the scale  characteristic of the Casimir
effect, is the marking difference between our scheme and previous
attempts at a direct measurement of the relative variation of the
Casimir force,  caused by a change in the reflective power of the
mirrors. For example, Ref. \cite{iann} reports on an experiment,
based on the technology of hydrogen switchable mirrors. Despite
the large modulation of reflectivity in the optical region of the
spectrum, which is very relevant for typical submicron separations
between the mirrors, no detectable variation in the Casimir force
could be seen, possibly due to lack of the necessary sensitivity
in force measurements.

Another comment on Eq. (\ref{shift})  is in order. As we shall see
in Sec. IV, and what is a distinctive feature of the Casimir
effect, the variation of Casimir energy $\Delta F_E^{(C)}$ turns
out to depend on $L$, while the condensation energy does not. This
implies an $L$ dependence of the shift of critical field, which,
if seen, would be  strong evidence for the Casimir origin of the
effect. In this respect, we notice that having chosen for the
outer plates a non superconducting material, we do not have to
worry about possible $L$-dependent Josephson couplings among the
film and the lateral plates, that could produce a similar effect.

\section{Calculation of the variation of Casimir energy in
the superconducting transition}

In this Section, we show how to compute the   difference  among
the Casimir free energies of the cavity, in the superconducting
($s$) and in the normal ($n$) states of the film (in zero magnetic
field).  In this computation we shall use the generalization to
multilayer systems of the theory developed by Lifshitz \cite{lifs}
to compute the van der Waals and Casimir energy for a
plane-parallel cavity formed by two thick dielectric walls
separated by an empty gap.

Before we turn to this computation, it is however necessary to
discuss first the applicability of the Lifshitz theory to a
superconducting cavity. Another concern is to  ascertain  whether
we can use that theory to reliably evaluate $\Delta F^{(C)}_E(T)$
for separations $L$ and for film thicknesses $D$ as small as a few
nm, which are the values of interest for us. Now, the main
assumption of Lifshitz theory is that, {\it in the relevant range
of frequencies and wave vectors}, one can describe the propagation
of electromagnetic waves inside the media forming the cavity, in
terms of a complex permittivity, depending only on the frequency
$\omega$. It is important to stress that the theory includes also
non-retarded effects, and hence it has as limiting cases both van
der Waals forces (that become important at small distances, like
those we consider) and Casimir forces. On this ground, it has been
used recently to study van der Waals interactions among thin metal
films (of thickness about 10 \AA), till very small separations (a
few \AA) \cite{bostrom}.

When spatial non-locality effects become important, so that the
complex permittivity becomes a function also of the wave vector
$\bf q$, the Lifshitz theory is no longer applicable
\cite{barash}(see also the discussion on p.126 in Ref.
\cite{bordag}).  Now it is clear that non-local effects are
important, in general, in superconductors, by virtue of the very
small skin depth (of the order of the penetration depth $\lambda$,
typically $500$ $\rm{\AA}$  or so) also in the frequency region
characteristic of the normal skin effect in normal metals (for an
interesting discussion of non-local effects in the computation of
dispersion forces in superconductors, see Ref. \cite{blos}).
Moreover, for small separations $L$, as well as for small film
thicknesses $D < 25-30$ nm and/or at cryogenic temperatures
non-local effects become important also in the normal state
\cite{svetov}, and indeed it has been advocated that space
dispersion should be taken into account, for example, to evaluate
the influence of thin metal coatings, that are used to protect the
plates in most of the current experiments on the Casimir effect
(see \cite{bordag} and last of Refs.\cite{svetov}. The influence
of space dispersion on the Casimir effect at room temperature is
studied also in the recent paper \cite{sern}). These
considerations lead to the conclusion that, for a reliable
evaluation of the {\it individual } Casimir free energies
$F^{(C)}_n(T)$ and $F^{(C)}_s(T)$, it is necessary to consider
space dispersion. However, and this is a key point, space
dispersion is unimportant for the purpose of computing the {\it
difference} $\Delta F^{(C)}_E(T)$ between the Casimir energies in
the two states of the film, which is the only quantity of interest
for us. The reason is that the optical properties of thin films
(with a thickness $D$ much smaller than the skin depth or
correlation length $\xi$), in the normal and in the
superconducting states, are indistinguishable for photon energies
larger than a few times $k \,T_c$, as accurate measurements have
shown \cite{glover}. This implies that, in the computation of
$\Delta F^{(C)}_E$, the only relevant photon energies are those
below a few times $k \,T_c$ (corresponding to the far IR), which
is where the optical properties of the film actually change when
it becomes superconducting. In this wavelength region, the quoted
experiment shows that the transmittivity data for thin
superconducting films can be well interpreted in terms of a
complex permittivity that depends only on the frequency, and is
independent of the film thickness.

Having established the applicability of the Lifshitz theory to our
superconducting cavity, we recall that in the original treatment
by Lifshitz, dealing with two thick dielectric slabs separated by
an empty gap, the Casimir force was obtained by evaluating, at
points in the gap region, the Maxwell stress tensor associated
with the electromagnetic fields generated by the randomly
fluctuating currents in the interior of the dielectric slabs.
Although physically transparent, this approach has never been
extended to multilayer systems, due to the complexity of the
computations that are involved. The desired generalization can
however be obtained by using an alternative approach, similar to
the one used by Casimir himself in his study of an ideal metallic
cavity, in which the Casimir energy is obtained as the zero-point
energy of the electromagnetic field inside the cavity. This method
was applied to dielectric cavities in Refs. \cite{kamp}, where it
was shown that one recovers the Lifshitz result, and was later
generalized to multilayer systems in Ref. \cite{zhou}.

We thus consider the five-layer system depicted in Fig. 1. The
electric permittivities of the layers are denoted as follows:
$\epsilon_{n/s}$ represents the permittivity of the film, in the
$n/s$ states respectively, while $\epsilon_1$ is the permittivity
of the insulating layers.  Last, $\epsilon_2$ is the permittivity
of the outermost thick normal metallic plates.

We consider first the $T=0$ case. Then, the Casimir free energy
coincides with the Casimir energy $E^{(C)}$, and we can write the
unrenormalized variation of Casimir energy $\Delta E^{(C)}_0(L,D)$
as
\begin{widetext} \be \Delta E^{(C)}_0(L,D)=A\,\frac{\hbar}{2}  \int
\frac{dk_1 dk_2}{(2 \pi)^2} \left\{\sum_p (\omega_{{\bf
k_\bot},\,p}^{(n,\,TM)}+\omega_{{\bf k_\bot},\,p}^{(n,\,
TE)})-\sum_p (\omega_{ {\bf k_\bot},\,p}^{(s,\,TM)}+\omega_{ {\bf
k_\bot},\,p}^{(s,\,TE)}) \right\}\;,\label{unren}\ee
\end{widetext}
where $A \gg L^2$ is the area of the cavity, ${\bf
k_\bot}=(k_1,k_2)$ denotes the two-dimensional wave vector in the
$xy$ plane, while $\omega_{{\bf k_\bot},\,p}^{(n/s,\,TM)}$
($\omega_{{\bf k_\bot},\,p}^{(n/s,\,TE)}$) denote the proper
frequencies of the TM (TE) modes,  in the $n/s$ states of the
film, respectively.

Upon using the  argument theorem, and by subtracting the
contribution corresponding to infinite separation $L$ (for
details, we address the reader to Chap. 4 in Ref. \cite{bordag})
\footnote{When comparing the formulae of this paper with those of
\cite{bordag}, please note that our $L$ and $D$ correspond,
respectively, to $d$ and $a$ of \cite{bordag}, while the TM and TE
modes are labelled there by the suffices (1) and (2),
respectively. Note also that in our configuration the central
layer is constituted by the superconducting film, and not by the
vacuum, and then its permittivity, denoted by $\epsilon_0$ in
\cite{bordag}, is not equal to 1, but rather to $\epsilon_{n/s}$
depending on the state of the film.}, we can rewrite   the
renormalized  sums   in Eq. (\ref{unren}) as integrals over {\it
complex} frequencies $i \zeta$:
\begin{widetext}
\be \left(\sum_p \omega_{{\bf k_\bot},\,p}^{(n,\,TM)} -\sum_p
\omega_{ {\bf k_\bot},\,p}^{(s,\,TM)}\right)_{\rm ren}=\frac{1}{2
\pi}\int_{-\infty}^{\infty} d \zeta\,  \left(\log
\frac{\Delta^{(1)}_n(i
\zeta)}{\widetilde{\Delta}^{(1)}_{n\,\infty}(i \zeta)}-\log
\frac{\Delta^{(1)}_s(i
\zeta)}{\widetilde{\Delta}^{(1)}_{s\,\infty}(i
\zeta)}\right)\;,\label{rensum}\ee
\end{widetext}
where ${\Delta}^{(1)}_{n/s}(i \zeta)$ is the expression in Eq.
(4.7) of \cite{bordag} (evaluated for $\epsilon_0=\epsilon_{n/s}$)
and $\widetilde{\Delta}^{(1)}_{n/s\,\infty}(i \zeta)$ denotes the
asymptotic value of ${\Delta}^{(1)}_{n/s}(i \zeta)$ in the limit
$L \rightarrow \infty$ (corresponding to the limit $d \rightarrow
\infty$ with the notation of \cite{bordag}). A similar expression
can be written for the $TE$ modes, which involves the quantity
${\Delta}^{(2)}_{n/s}(i \zeta)$ defined in Eq. (4.9) of
\cite{bordag}. Upon inserting Eq. (\ref{rensum}), and the
analogous expression for  $TE$ modes, into Eq. (\ref{unren}) one
gets the following expression for the (renormalized) variation
$\Delta E^{(C)}(L,D)$ of the Casimir energy:
\begin{widetext}\be
 \Delta E^{(C)}= A \;\frac{\hbar}{2}\int   \, \frac{d {\bf k_{\bot}}}{(2 \pi)^2} \int_{-\infty}^{\infty} \frac{d
\zeta}{2 \pi}\, \,\left(\log \frac{Q_n^{TE}}{Q_s^{TE}}+\log
\frac{Q_n^{TM}}{Q_s^{TM}}\right)\;, \label{lif}\ee
\end{widetext} where we set \be Q^{(TM/TE)}_I(\zeta) \equiv
\frac{\Delta^{(1/2)}_I(i
\zeta)}{\widetilde{\Delta}^{(1/2)}_{I\,\infty}(i
\zeta)}\;,\;\;\;I=n,s\,. \ee Upon performing the change of
variables $k_\bot^2=(p^2-1)\zeta^2/c^2$ in the integral over $
k_\bot $,  the above expression for  $\Delta E^{(C)}(L,D)$  turns
into
\begin{widetext} \be \Delta E^{(C)}=\frac{\hbar A}{4 \pi^2
c^2}\int_1^{\infty} p\,dp \int_0^{\infty} d \zeta\,\zeta^2 \,
\left(\log \frac{Q_n^{TE}}{Q_s^{TE}}+\log
\frac{Q_n^{TM}}{Q_s^{TM}}\right)\;, \label{denren}\ee
\end{widetext} where the explicit expression of the coefficients
$Q^{(TM/TE)}_I$ is
\begin{widetext}
\begin{eqnarray}
 Q_I^{TE/TM}( \zeta,p)=
\frac{(1-\Delta_{1I}^{TE/TM}\Delta_{12}^{TE/TM}e^{-2 \zeta\,K_1  \,
L/c})^2 -(\Delta_{1I}^{TE/TM}-\Delta_{12}^{TE/TM}e^{-2 \zeta \,K_1\,
L/c})^2 e^{-2 \zeta K_I D/c}}{1-(\Delta_{1I}^{TE/TM})^2 e^{-2
\zeta K_I \,D/c}}\;,\label{delec}\\
\Delta_{j\,l}^{TE}=\frac{K_j- K_l}{K_j+K_l}\;,\;\;
\Delta_{j\,l}^{TM}=\frac{K_j \,\epsilon_l\,(i \zeta)-K_l
\,\epsilon_j\,(i \zeta)}{K_j\, \epsilon_l\,(i \zeta)+K_l\,
\epsilon_j\,(i \zeta)}\;,\;\;K_j=\sqrt{\epsilon_j\,(i
\zeta)-1+p^2}\;,\;\;\;I=n,s\;\;;\;\;j\,,\,l=1,2,n,s.\label{defs}
\end{eqnarray}
\end{widetext}

Formulae along similar lines, involving the reflection
coefficients of the layers, have been obtained for example in
Refs.\cite{lambr}.

The extension of the above formulae to the case of finite
temperature is straightforward. As is well known this amounts to
the replacement in Eq. (\ref{lif}) of the integration $ \int d
\zeta/2 \pi$ by the summation $k T/\hbar \sum_l$ over the
Matsubara frequencies $\zeta_l=2 \pi l/\beta$, where
$\beta=\hbar/(k T)$, which leads to the following expression for
the variation $\Delta F^{(C)}_E(T)$ of  Casimir free energy:
\begin{widetext}\be \Delta F^{(C)}_E(T)=A\, \frac{k
\,T}{2}\sum_{l=-\infty}^{\infty}\int \frac{ d {\bf k_\bot}}{(2
\pi)^2} \,\left(\log \frac{Q_n^{TE}}{Q_s^{TE}}+\log
\frac{Q_n^{TM}}{Q_s^{TM}}\right)\;. \label{fint}\ee
\end{widetext}
As we see, Eqs. (\ref{denren}-\ref{fint}) involve the electric
permittivities $\epsilon\,(i \zeta)$ of the various layers at
imaginary frequencies $i \zeta$. For these functions, we have made
the following choices.

For the outermost metal plates, we use  a Drude model for the
electric permittivity: \be
\epsilon_D(\omega)=1-\frac{\Omega^2}{\omega(\omega+i
\gamma)}\;,\label{drper}\ee where $\Omega$ is the plasma frequency
and $\gamma=1/\tau$, with $\tau$ the relaxation time. We denote by
$\Omega_2$ and $\tau_2$ the values of these quantities for the
outer plates. As is well known, the Drude model provides a very
good approximation in the low-frequency range $\omega \approx 2
k\, T_c/\hbar \simeq 10^{11}\div 10^{12}$ rad/sec which is
involved in the computation of $\Delta F^{(C)}_E(T)$ and $\Delta
E^{(C)}$. The relaxation time is temperature dependent and for an
ideal metal it becomes infinite at $T=0$. However, in real metals,
the relaxation time stops increasing  at sufficiently low
temperatures (typically of the order of a few K), where it reaches
a saturation value, which is determined by the impurities that are
present in the metal. Since  in a superconducting cavity the
temperatures are very low, we can assume that $\tau_2$ has reached
its saturation value and therefore we can treat it as a constant.
The  continuation of Eq. (\ref{drper}) to the imaginary axis is of
course straightforward and gives \be \epsilon_D(i
\zeta)=1+\frac{\Omega^2}{\zeta\,(\zeta+ \gamma)}
\;.\label{druima}\ee

For the insulating layers, we take a constant permittivity, equal
to the static value:
 \be
\epsilon_1(\omega)=\epsilon_1(0)\;.\ee Again, this is a good
approximation in the range of frequencies that we consider.

As for the film, in the normal state we use  again the Drude
expression Eq. (\ref{drper}),  with appropriate values for the
plasma frequency $\Omega_n$ and the relaxation time $\tau_n$.

The permittivity $\epsilon_s(i \zeta)$ of the film in the
superconducting state cannot be given in closed form and we have
evaluated it as explained in Appendix B, starting from the
Mattis--Bardeen  formula for the conductivity $\sigma_s(\omega)$
of a superconductor in the local limit $q \rightarrow 0$ of BCS
theory.

\section{Numerical computations}

In this Section we present the results of our numerical
computations. For the convenience of the reader, the material is
divided into four separate Subsections. In the first Subsection we
analyze the qualitative features to be expected from a numerical
evaluation of Eqs. (\ref{lif})--(\ref{fint}). In Subsec. B we
present the computation of the variation $\Delta F^{(C)}_E$ of
Casimir free energy in the superconducting transition. In Subsect.
C we discuss the controversial issue of the contribution arising
from the $TE$ zero mode. Finally, in the last Subsection, we
present our results for the shift of critical magnetic field.

\subsection{Qualitative features of $\Delta F^{(C)}_E$}

In this Subsection we discuss some qualitative features of Eqs.
(\ref{lif})--(\ref{fint}) that are useful as a check of the
numerical computations. We recall first of all that, with our
choices for the cavity geometry and for the permittivities of the
various layers, the value of $\Delta F^{(C)}_E$ depends on the
following sets of parameters:
\begin{enumerate}
     \item the thickness of the film $D$ and the width of the insulating gaps $L$ (see Fig.
    1);
    \item the temperature $T$, which we shall express in units of
    $T_c$, by means of $t=T/T_c$;
    \item the plasma frequency $\Omega_2$ and the saturation value
    $\tau_2$ for
    the relaxation time
    of the outer plates;
    \item the dielectric constant $\epsilon_1(0)$ of the insulating
    gaps;
    \item the  plasma frequency $\Omega_n$ and the saturation
    value $\tau_n$ for the relaxation time
     of the film, in its normal state;
    \item the BCS gap $\Delta (T)$ of the superconducting film.
\end{enumerate}
Some simple properties of our expression for $\Delta F^{(C)}_E(T)$
are obvious. First, it is clear from Eq. (\ref{delec}) that $
Q_I^{TE/TM}( \zeta,p) \rightarrow 1$ for $L \rightarrow \infty$,
and thus $\Delta F^{(C)}_E  \rightarrow 0$ in this limit, as it
should. Another important limit is that for $L \rightarrow 0$. In
this limit, $\Delta F^{(C)}_E$ approaches a {\it finite} value,
because, as pointed earlier, for large $\zeta$ the permittivities
of the film in the normal and in the superconducting states become
undistinguishable, and then the ratios
${Q_n^{TE/TM}}/{Q_s^{TE/TM}}$ approach one (and their logarithms
approach zero)   faster than $Q_{n/s}^{TE/TM}$,  and this ensures
convergence of Eqs. (\ref{denren}) and (\ref{fint}) also for
$L=0$.

We  consider now the dependence of $\Delta F^{(C)}_E(T)$ on $D$.
For $D \rightarrow \infty$, the expressions of $Q_I^{TE/TM}$
reduce to \be \lim_{D \rightarrow \infty} Q_I^{TE/TM}=
(1-\Delta_{1I}^{TE/TM}\Delta_{12}^{TE/TM}e^{-2 \zeta\,K_1 \,
L/c})^2\;,\label{limqd} \ee which concides with the square of the
analogous expression for a three-layer cavity consisting only of
the outer plate, the insulating layer and an infinite
superconducting layer.  Upon taking the logarithm of $Q_I^{TE/TM}$
we thus obtain that $\Delta F^{(C)}_E$ is twice the value   for
the three-layer system, which is what one would have expected
since in the limit $D \rightarrow \infty$ our five-layered cavity
becomes equivalent to a system of two identical three-layered
cavities.  The rate of approach to the limiting double three-layer
system is controlled by the exponential factors $\exp (-2 D
K_{n/s} \zeta /c)$, which set the scale $D_0$  above which $\Delta
F^{(C)}_E$ becomes independent of $D$ and reaches its $D
\rightarrow \infty$ limiting value. The value of $D_0$ can be
estimated as follows.  For photon frequencies $\zeta$ of order
$\bar{\zeta}= 2 \Delta /\hbar$, which are the  relevant ones for
${\Delta F^{(C)}_E}$, it turns out that $\epsilon_{s}(i \zeta)
> \epsilon_{n}(i \zeta)$ (see Appendix B), and therefore we have
that $K_{s}(\zeta,p)
> K_{n}(\zeta,p)>K_{n}(\zeta,0)$. On the other hand, using the Drude
model Eq. (\ref{druima})  for $\epsilon_{n}(i \zeta)$  we estimate
$\zeta\, K_{n}(\zeta,0) \approx \zeta \,\sqrt{\epsilon_{n}(i\zeta)}
\approx   \Omega_n/{\sqrt{1+y}}$, which implies
$$\exp({-2 D K_{n/s} \zeta
/c}) < \exp \left(-\; \frac{2\,D\,      \Omega_n}{ \,c\,
\sqrt{1+y}}\right)\;.$$ We see from this Equation that the
exponential becomes negligible for $D \gg D_0$, where \be
D_0=\frac{c}{2\,\Omega_n}\,\sqrt{1+y}\;.\ee For a typical value of
the plasma frequency $\hbar \Omega_n \simeq (10 \div 20)$ eV,  we
thus obtain
$$
D_0 \approx 10\; {\rm nm}\;.
$$
The above number is very important for us, because it provides an
estimate of the thickness $D$ of the film, for which the largest
shift of critical field should be expected. Indeed, we see from
Eq. (\ref{shift}) that the largest shift is obtained  for the
value of $D$ that maximizes the ratio ${\Delta F^{(C)}_E}/{{\cal
E}_{\rm cond} }$. Now, the condensation energy ${{\cal E}_{\rm
cond} }=V \, e(T)$ is proportional to the volume $V$ of the film
and then to $D$ (in reality, for certain superconductors and for
ultrathin films with $D \ll \lambda, \xi$ the critical temperature
$T_c$ and then the density of condensation energy $e(t)$ may
depend on $D$. See discussion on Be films in the next Subsection).
Therefore, for $D \gg D_0$, when ${\Delta F^{(C)}_E}$ becomes
independent of $D$, the ratio ${\Delta F^{(C)}_E}/{{\cal E}_{\rm
cond} }$ decreases like $1/D$  because of  the unlimited increase
of ${\cal E}_{\rm cond}$ with $D$.  It then follows that the
optimal values of $D$ must be of order $D_0$ or so, i.e. of a few
nm, and this is the range that we shall consider in what follows.

\subsection{Numerical computation of the variation of Casimir
free energy}

The  variation of Casimir free energy $\Delta F^{(C)}_E(T)$, Eq.
(\ref{fint}),  has been computed numerically. For a
superconducting film with a critical temperature $T_c$, the gap
$\Delta(T)$ has been estimated by using the approximate BCS
formula Eq. (\ref{gap}). In what follows we present some of our
results in terms of the so-called impurity parameter $y=\hbar/(2
\Delta(T) \tau_n)$, which is commonly used, in place of the
relaxation time $\tau_n$, as a convenient measure of the degree of
purity of the superconductor. In the range of parameters that we
consider, the value of $\Delta F^{(C)}_E$ is independent,   to
better than four significant digits, of the value of
$\epsilon_1(0)$, for $1<\epsilon_1(0)<300$. In all  computations
presented below, we fix once and for all $\epsilon_1(0)=100$.

Another important remark is that, in all cases that we considered,
the contribution of $TM$ modes to $\Delta F^{(C)}_E$ is completely
negligible, being three or four orders of magnitude smaller than
that of $TE$ modes. We postpone to next Section a discussion of
the implications of this fact.

In Fig. 2 we show the plot of $\Delta F^{(C)}_E$ (in erg) as a
function of the width $L$ (in nm) of the insulating gap, for $D=5$
nm, $T_c=0.5$ K, $y=15$, $t=0.9$, $\Omega_n=\Omega_2=18.9$ eV,
$\tau_2=2.4 \times 10^{-12}$ sec.   We observe that $\Delta
F^{(C)}_E$ is always {\it positive}, which corresponds to the
intuitive expectation that transition to superconductivity of the
film leads to a {\it stronger} Casimir effect, i.e. to {\it lower}
Casimir free energy. The data can be fit very accurately by a
curve of the type \be \Delta F^{(C)}_E(L)  \propto \frac{1}{1+
(L/L_0)^{\alpha}}\;,\ee where $L_0=8.3$ nm and $\alpha=1.15$. We
thus see, as anticipated earlier, that $\Delta F^{(C)}_E(L)$
approaches a finite limit for $L \rightarrow 0$.
\begin{figure}
\includegraphics{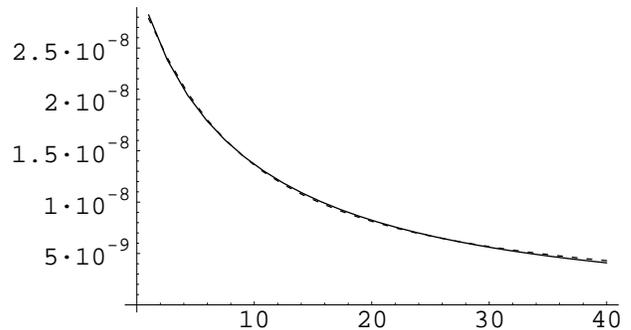}
\caption{\label{figlungo1} Plots of   $\Delta F^{(C)}_E$ (in erg)
as a function of $L$ (in nm)  for $D=5$ nm, $T_c=0.5$ K, $t=0.9$.
See text for the values of the other parameters. Also shown is the
plot (dashed line) of a fit of the type $1/(1+(L/L_0)^{\alpha})$,
with $L_0=8.3$ nm and $\alpha=1.15$.}
\end{figure}
This  feature of $\Delta F^{(C)}_E$ is helpful, from an
experimental point of view, because it implies that the possible
roughness of the film and  of the lateral plates will have little
influence on the value of $\Delta F^{(C)}_E$, for $L$ sufficiently
small. By contrast, we recall that the roughness of the surfaces
is one of the main sources of uncertainty in the theoretical
analysis of standard measurements of the Casimir force, for
submicron separations \cite{bordag}.

In Fig. 3 we show the plots of $\Delta F^{(C)}_E$ (in erg), as a
function of the  impurity parameter $y$, for the following three
combinations of values of the plasma frequencies for the  film and
the lateral plates: $\Omega_n=\Omega_2=18.9$ eV (solid curve),
$\Omega_n=18.9$, $\Omega_2=12$ eV (point-dashed curve) and
$\Omega_n=\Omega_2=12$ eV (dotted curve) (the values of 18.9  eV
and 12 eV  correspond respectively, to  Be and Al). All curves
have been computed for $L=10$ nm, $D=5$ nm, $T_c=0.5$ K, $t=0.9$,
$\tau_2=2.4 \times 10^{-12}$ sec.   We see   that $\Delta
F^{(C)}_E$ has a maximum for $y \simeq 10 \div 15$. While the
decrease of $\Delta F^{(C)}_E$  for large values of $y$ is
obvious, it is perhaps surprising to see that it decreases also
when $y$ becomes very small, i.e. when the degree of purity of the
film is improved. In fact, this latter behavior is explained by
observing that, for small values of $y$, the film is so pure that
already in the normal state, it behaves as a very good conductor
and then transition to superconductivity cannot increase
significantly the Casimir energy. Notice also that $\Delta
F^{(C)}_E$ is   very sensitive to the values for the plasma
frequencies both of the film and of the lateral plates, and it
increases sensibly when either one is increased.

\begin{figure}
\includegraphics{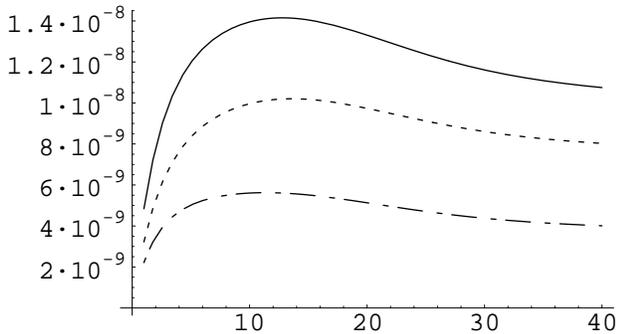}
\caption{\label{figlungo2} Plots of   $\Delta F^{(C)}_E$ (in erg)
for $L=10$ nm, $D=5$ nm,  $T_c=0.5$ K and $t=0.9$,  as a function
of the impurity parameter $y$, for $\Omega_n=\Omega_2=18.9$ eV
(solid line), $\Omega_n=18.9$, $\Omega_2=12$ eV (dashed line) and
for $\Omega_n=\Omega_2=12$ eV (point-dashed line). See text for
values of the other parameters.}
\end{figure}
In Fig. 4 we show the plot of $\Delta F^{(C)}_E$ as a function of
the critical temperature $T_c$ of the film, for $L=10$ nm, $D$=5
nm, and $t=0.9$. All other parameters are  as in Fig. 2. The
figure shows that $\Delta F^{(C)}_E$ depends linearly on $T_c$.

\begin{figure}
\includegraphics{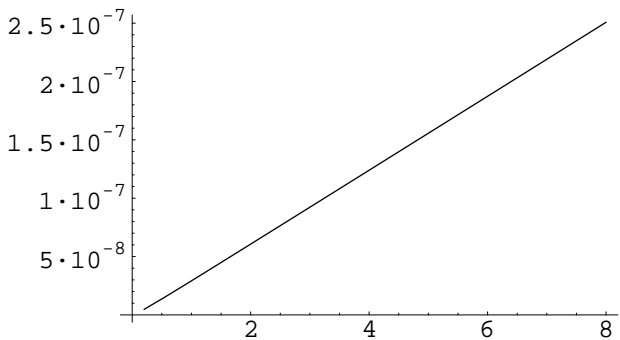}
\caption{\label{figlungo3} Plots of   $\Delta F^{(C)}_E$ (in erg)
as a function of $T_c$ (in K) for $L=10$ nm, $D=5$ nm,  $y=15$ and
$t=0.9$. See text for  values of the other parameters.}
\end{figure}
In Fig. 5 we show the plot of $\Delta F^{(C)}_E$ as a function of
$t$, for $T_c=0.5$  and $\tau_n=5 \times 10^{-13}$ sec (all other
parameters are   as in Fig. 4).  Notice that in this plot we are
holding $\tau_n$ constant and hence, by virtue of the temperature
dependence of the gap $\Delta(T)$ (see Eq. (\ref{gap})),  the
impurity parameter $y$ is not constant (for $T=0$, we have
$y(0)=8.7$).  Note also that, for $T/T_c \rightarrow 1$, $\Delta
F^{(C)}_E$ approaches zero linearly in $(1-t)$. Also shown in the
Figure (dashed line) is the plot of the low-temperature
approximation to $\Delta F^{(C)}_E$, obtained by replacing the
Matsubara sum Eq. (\ref{fint}) by the integral over $\zeta$, as in
Eq. (\ref{denren}). This approximation was used in Ref.
\cite{bimonte}, and as we see from Fig. 5 it works fairly well in
the  whole range of temperatures. For the explanation of the third
curve (point-dashed line) shown in Fig. 5, see next Subsection.

\subsection{Contribution from the $TE$ zero mode.}

In the current literature on thermal corrections to the Casimir
effect there is an ongoing controversy concerning the proper way
of calculating the contribution of the $TE$ zero mode (i.e. the
$l=0$ term  in  the Matsubara sum) to the Casimir free energy (see
Refs. \cite{geyer} for a detailed discussion of different points
of view on this problem). The essence of the question is whether
one can use the Lifshitz theory to evaluate the $TE$ zero mode,
and in the affirmative case, for what choice of the metal
permittivity function $\epsilon(i \zeta)$. Indeed, all the trouble
arises from the fact that the computation of this mode involves
the quantity $C$: \be C:=\lim_{\zeta \rightarrow 0} (\zeta^2
\,\epsilon(i \zeta)) \;.\ee If one uses the Drude function Eq.
(\ref{druima}) (with a finite value of $\tau$) one obtains $C=0$,
and this implies that the $TE$ zero  mode gives zero contribution
to the Casimir free energy, irrespective of how large $\tau$ is.
On the contrary, if one uses the simpler plasma model that gives
\be \epsilon(i \zeta)=1+\frac{\Omega_p^2}{\zeta^2}\;\ee one finds
$C=\Omega_p^2$, and then the zero mode gives a non vanishing
contribution, reproducing the ideal metal case in the limit of
infinite plasma frequency. While we address the reader to Refs.
\cite{geyer} for a discussion of the reasons in favor of the
alternative approaches to this question that have been proposed by
a number of authors, we would like to examine the consequences of
this problem for our computation of $\Delta F^{(C)}_E$.
\begin{figure}
\includegraphics{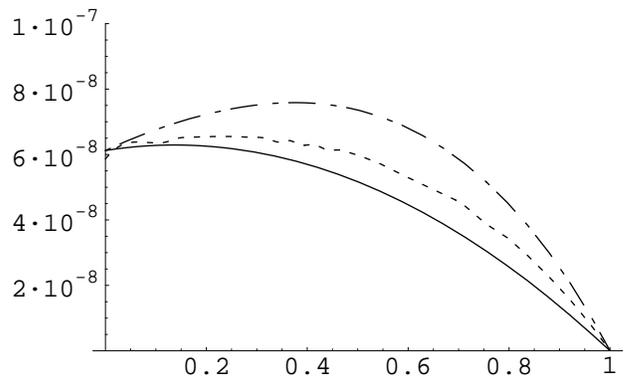}
\caption{\label{figlungo5} Plot (solid line) of   $\Delta
F^{(C)}_E$ (in erg) as a function of $t$  for $L=10$ nm, $D=5$ nm,
$T_c=0.5$ K, $\tau_n=5 \times 10^{-13}$ sec. The point-dashed line
includes the contribution of the $TE$ zero mode, computed using
the plasma model for the lateral plates. Also shown (dashed line)
is the plot of the low-temperature limit of the Matsubara sum, Eq.
(\ref{denren}). See text for further details.}
\end{figure}

All computations presented in the previous Section used the Drude
model for the permittivities for both  the lateral plates and the
film, in the normal state. Since, as pointed out earlier, the
contribution from the $TM$ modes is negligible, it follows that
the values of $\Delta F^{(C)}_E$ that we computed receive no
contribution from the $TE$ zero mode.

It is interesting to see what happens in our computations if, in
the evaluation of the zero mode, we replace the Drude model by the
plasma model. We shall do it only for the lateral plates. For the
superconducting film we prefer keep using the Drude model, since
it is the one to which the BCS formula of the permittivity
converges for $T \rightarrow T_c$. Thus, if we denote by $\Delta
F^{(C)}_E(TE \;{\rm z.m.})$ the contribution of the $TE$ zero
mode, we obtain for the total  variation of Casimir free energy
the expression \be \widetilde{\Delta F}^{(C)}_E =\Delta
F^{(C)}_E(TE \;{\rm z.m.})+\Delta F^{(C)}_E \;. \ee

In Fig. 5 we plot (dashed line)  $\widetilde{\Delta F}^{(C)}_E$ as
a function of $t$, while in Fig. 6 we  show the  ratio $\Delta
F^{(C)}_E(TE \;{\rm z.m.})/\Delta F^{(C)}_E$, again as a function
of $t$ (all parameters are   as in Fig. 5). We observe that the
weight of the zero mode increases as we approach $T_c$. The reason
of this is easy to understand: the   zero mode becomes more and
more important because  a decreasing number of Matsubara modes
contribute to $\Delta F^{(C)}_E$, as one approaches $T_c$. This is
so because the quantities $(Q_{n}^{TE}/Q_{s}^{TE})(i \zeta)$ that
occur in Eq. (\ref{fint}) are substantially different from one
only for complex frequencies $\zeta$ of  order of a few tens of
times $k \,T_c/\hbar$. Since the $l$-th Matsubara mode has a
frequency equal to $2 \pi \,l\,k\, T/\hbar$, it is clear that the
number of terms effectively contributing to $\Delta F^{(C)}_E$
should be roughly proportional to $T_c/T$, and hence it is large
for $T \ll T_c$, but becomes small for $T$ comparable to $T_c$.
Notice that close to $T_c$ the contribution of the zero mode is
practically of the same magnitude as that of $\Delta F^{(C)}_E$
and therefore its inclusion  doubles the variation of Casimir free
energy.

\subsection{Shift of critical field}

Having estimated the variation of Casimir energy $\Delta
F^{(C)}_E$ across the transition, we are now ready to compute the
shift of parallel critical field, using Eq. (\ref{hcricav}). The
result depends of course on the superconductor, and in order to
get a feeling of what to expect depending on this choice,  it is
convenient to consider the approximate expression for the relative
shift Eq. (\ref{shift}). According to this formula, the shift is
larger when ${\cal E}_{\rm cond}$ is smaller and/or $\Delta
F^{(C)}_E$ is larger.  Now, we see from Fig. 4 that $\Delta
F^{(C)}_E$ is  almost proportional to $T_c$, while from Eqs.
(\ref{econtd}) and (\ref{ezero}) in Appendix A  we see $ {\cal
E}_{\rm cond}(T) \propto  H_c^2(0)$. Therefore, if we  forget for
a moment     other influential factors (like the values for the
impurity parameter $y$, the plasma frequency and the relaxation
time that we shall consider later) we obtain from Eq.
(\ref{shift}) the estimate \be \frac{\delta H_{c \|}}{H_{c \|}}
\approx \frac{\Delta F^{(C)}_E}{2\,{\cal E}_{\rm cond}(T)} \propto
\frac{T_c}{ H_c^2(0)}\;.\ee  We thus see that the relative shift
should be inversely proportional to the quantity
$s:=H_c^2(0)/T_c$, and so superconductors with small values of $s$
should be preferable. Now, if we use the empirical law Eq.
(\ref{emptc}) to estimate the dependence of $H_c(0)$ on the
critical temperature of the superconductor, we find that $s$ goes
like $T_c^{1.6}$, and this implies that we should  first consider
superconductors with low $T_c$. In reality, things are not so
simple.
\begin{figure}
\includegraphics{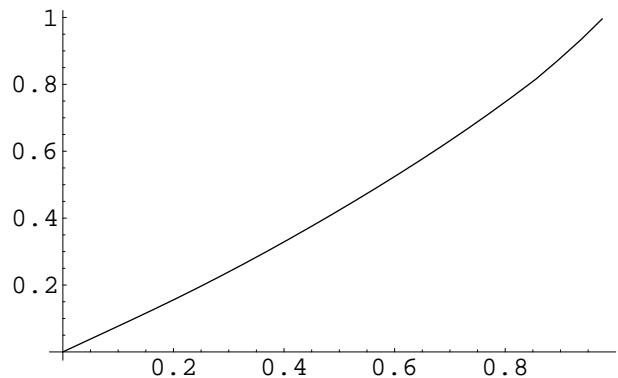}
\caption{\label{figlungo5} Plot of  the  ratio $\Delta
F^{(C)}_E(TE \;{\rm  z.m.})/\Delta F^{(C)}_E$ as a function of
$t$.  All parameters are same as in Fig. 5. $\Delta F^{(C)}_E(TE
\;\rm {z.m.})$ is evaluated using the plasma model for the outer
plates.}
\end{figure}
Consider for example the case of Be. It has a very low $T_c=24$
mK, and a correspondingly low $H_c(0)=1.079 \pm 0.002$ Oe
\cite{soulen},  giving $s=48.5$ Oe$^2$/K, which is much smaller
than the values of $s$ for other superconductors. As a comparison,
we   quote in Table I the values of $T_c$, $H_c(0)$ and $s$ for a
number of type I superconductors.
\begin{table}
\caption{\label{tab:table1}  Values   of $T_c$ (in K), $H_c(0)$
(in Oe) and $s=H_c^2(0)/T_c$ in (Oe$^2$/K) for some type I
superconductors.}
\begin{ruledtabular}
\begin{tabular}{cccc}
 &    $T_c$  &   $H_c(0)$   &   $H_c^2(0)/T_c$  \\
\hline Be & 0.024 & 1.08 & 48.5 \\
Zn  &  0.88   &   53 &  3192  \\
Al  & 1.2  & 105  & 9343   \\
Sn  & 3.72  & 306 & 25171\\
\end{tabular}
\end{ruledtabular}
\end{table}
From these data it would seem that Be should give a great
advantage over other superconductors. However, unfortunately, the
above low value of $T_c$ refers to bulk samples, while  very thin
Be films (with thicknesses of a few nm) have a much higher $T_c$.
For example, the authors of Ref. \cite{adam2} report a $T_c$ of
0.5 K for a thickness of 5 nm. The value of the condensation
energy for such a film is not reported in Ref. \cite{adam2}, but
one can estimate that ${\cal E}_{\rm cond}$ scales with respect to
the bulk value in the same proportion as the square of the
critical temperature (see Appendix A). If we insist in expressing
the condensation energy in the form of Eq. (\ref{ezero}), this
amounts to saying that the thermodynamical field of a thin film
scales with its critical temperature.  This gives us an estimate
of $H_c(0)=$22.5 Oe for the films of Ref. \cite{adam2},  and thus
we find $s=1010$ Oe$^2$/K, which is much worse than the bulk
value, but still good compared with other type I superconductors.

We consider now the other factors  affecting the relative shift of
critical field, that we have neglected in the above discussion.
The first important factor is the value of the plasma frequency
$\Omega$. As we see from Fig. 3, larger values of $\Omega$ both
for the film and the lateral plates  lead to significantly larger
values for $\Delta F^{(C)}_E$.   Now we recall that Be has $\hbar
\Omega=18.9$ eV, which is much  larger than the value for other
metals (for example Al, which has $\hbar \Omega=$ 12 eV).
Therefore it would be advantageous to use this material both for
the film and the lateral plates. In this respect, the circumstance
that the critical temperature of a thin Be film is much higher
than the bulk value  can be exploited to our advantage, as it
allows to use Be both for the film and the lateral thick plates,
leaving us with a wide range of temperatures for which the film is
superconducting, while the lateral mirrors are normal, as we
require. According to Fig. 3, we can gain in this way a factor
larger than two, with respect to say, a cavity made solely of Al.
\begin{figure}
\includegraphics{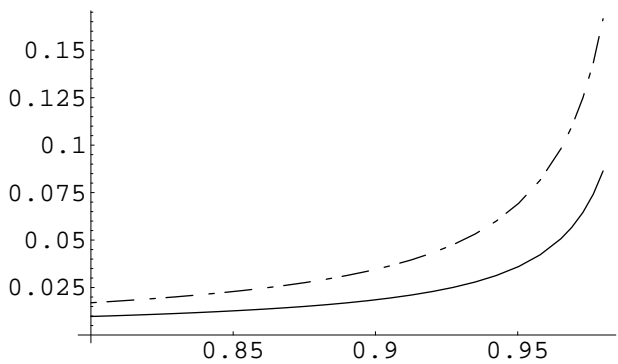}
\caption{\label{figlungo8} Plot (solid line) of  the relative
shift of parallel critical field for a Be cavity, as a function of
$T/T_c$. The point-dashed line includes the contribution of the
$TE$ zero modes, evaluated as in Fig. 5. All parameters are as in
Fig. 5.}
\end{figure}
Another reason in favor of Be is related to the fact that the
critical field $H_{c \|}$ of thin films is not defined in general
as sharply as in bulk samples, and one rather has a fuzzy critical
region of finite width. It is clear therefore that preference
should be given to materials for which this indeterminacy is
smallest. It turns out that Be can be deposited in ultrathin
films, characterized by an extraordinary degree of uniformity. A
consequence of this uniformity is that even for thicknesses $D$ of
a few nm, Be films possess very sharp parallel critical magnetic
fields $H_{c \|}$, that can be measured with a precision of a few
parts in a thousand \cite{adam2}.

Having  collected a number of arguments in favor of Be,  we can
now show  the results of our computations for the shift of
critical field. In Fig. 8  we show (solid line) the relative shift
of the critical parallel field of a Be film, as a function of $t$.
Note that, according to Eq. (\ref{shift}) this quantity is
independent of the coefficient $\rho$, whose precise value is in
general not known accurately. The Figure has been drawn using the
same parameters as in Fig. 5. We show also (point-dashed curve)
the relative shift that is obtained after including the
contribution of   $TE$ zero modes, computed in the way explained
in the previous Subsection. We see that this inclusion produces
almost a doubling of the shift.

Note that the shift is {\it positive}, meaning that the critical
field for the film placed in the cavity is {\it larger} than the
critical field for an identical film outside the cavity. The
increase of critical field relative shift as one approaches the
critical temperature  arises because, for $t \rightarrow 1$,
$\Delta F^{(C)}_E$ and ${\cal E}_{\rm cond}$ approach zero at
different rates. Indeed, while ${\cal E}_{\rm cond}$ vanishes as
$(1-t)^2$ (see Eq. (\ref{econtd})), we see from Fig. 5 that
$\Delta F^{(C)}_E$ vanishes essentially like the first power of
$1-t$.

\section{Conclusions and discussion}

We have studied a superconducting Casimir cavity, including a thin
superconducting film and we have shown that the variation of
Casimir free energy, accompanying the superconducting transition,
determines a measurable shift of the critical magnetic field
required to destroy the film superconductivity. This constitutes a
novel approach to the study of the Casimir effect, in which the
central role is played by the Casimir energy itself, rather than
the Casimir force, as in all experiments  performed so far.

Our scheme, apart from its novelty, presents a number of appealing
features. The first advantage stems from the fact that it uses
{\it rigid} cavities. As is well known, the experimental
difficulty of controlling the parallelism among macroscopic plane
plates with submicron separations led the experimenters to
consider simpler geometries that do not suffer from this problem,
like the sphere-plane one, which has been adopted in all precision
experiments on the Casimir force (with the only exception of the
experiment by Bressi et al. \cite{decca}, where the plane-parallel
configuration is used). This limitation has made it impossible so
far to explore experimentally one of the most intriguing features
of the Casimir effect, i.e. its dependence on the geometry of the
cavity.  The use of rigid cavities, as in our approach,  might
make it easier to realize  a number of geometries that are
difficult to realize within the standard force-measurements
scheme.

Another interesting feature of our scheme, not mentioned in Ref.
\cite{bimonte}, is its sensitivity to the contribution from the TE
zero mode. This is a controversial issue in the current literature
on thermal corrections to the Casimir effect. It is  well known
that, for submicron plate separations, thermal corrections  are
negligible at cryogenic temperature, and become relevant  only at
room temperature. However, this is not the case in our
superconducting cavities, because the shift of critical field is
completely determined by the Matsubara modes with frequencies
below or of order $k T_c/\hbar$, which is where the reflective
properties of a film change when it becomes superconducting. As a
consequence, as we have shown in Sec. IV C, different treatments
of the TE zero mode lead to strongly different predictions for the
shift of critical field, at the level of doubling the shift close
to $T_c$, and this opens the way to a possible experimental
clarification of this delicate problem.

Our computations show that larger shifts of the critical field can
be obtained by employing for the lateral plates and the
superconducting film  materials with high plasma frequencies,
which ensure a high reflection coefficient. We found also that it
is advantageous to use films with a value of the impurity
parameter $y$ about $10 \div 15$. Such values of $y$ require films
with rather large free mean paths, of  order 100 nm or so, which
can be difficult to reach  in ultrathin films with thicknesses of
$5 \div 10$ nm, as we require. Finally, since larger relative
shifts are found near the critical temperature, an effort should
be made to produce ultrathin films with high degree of purity, in
order to obtain narrow transition regions, with sharply defined
critical temperatures.

It is also the case to note that the variation of Casimir free
energy $\Delta F^{(C)}_E$ approaches a finite limit, for vanishing
separation $L$ of the film from the outer plates. For separations
of order 10 nm, this behavior of $\Delta F^{(C)}_E$ ensures a good
degree of robustness of the resulting shift of critical field with
respect to the roughness of the metal surfaces, which is known to
give an important contribution in standard Casimir force
measurements, for submicron separations \cite{bordag}. Equally
helpful in this respect is the fact that $\Delta F^{(C)}_E$ is
independent, to a high accuracy level, of the dielectric function
of the insulating layers separating the film from the outer
plates.

It would be very interesting to obtain an experimental
verification of the  phenomena described in this paper. Indeed
this is the aim of the ALADIN experiment, sponsored by INFN, which
is currently under way at the Dipartimento di Scienze Fisiche
dell'Universit\'a di Napoli Federico II. Some details about the
experimental setup are now helpful. In order  to detect the shift
in the critical field of a thin superconducting film resulting
from the Casimir effect, we foresee to simultaneously deposit on a
single substrate a certain number of superconducting films of same
thickness, covered with a thin insulating layer, again of same
thickness for all films. Half of these films will  then  be
covered with a third thick layer of a non-superconducting metal,
thus forming a three-layer Casimir cavity. Such three-layer
systems are easier to manufacture than the five-layer cavities
discussed in this paper, and thus, even if they lead to a smaller
effect, will be considered first. Passage to the five-layer case
is expected  at a later stage, and will be dealt with in a similar
way. The simultaneous deposition as above is aimed at ensuring
that all superconducting films, both the single ones and those in
the cavities, have the same properties (thickness, degree of
purity etc.). We plan to repeat the measurements for various film
and oxide thicknesses, in order to test our theoretical
predictions.

Besides the authors of the present paper, the experiment ALADIN
has involved  D. Born, A. Cassinese, F. Chiarella, L. Milano, O.
Scaldaferri, F. Tafuri, and R. Vaglio.

\begin{acknowledgments}

We would like to thank A. Cassinese, F. Tafuri, A. Tagliacozzo and
R. Vaglio for valuable discussions, G.L. Klimchitskaya and V.M.
Mostepanenko for enlightening discussions on several aspects of
the problem. G.B. and G.E. acknowledge partial financial support
by PRIN {\it SINTESI}.

\end{acknowledgments}

\section{Appendices}

For the convenience of the reader, we briefly review in these
Appendices the basic properties of type I superconductors, that
are important for the superconducting Casimir cavities to be
considered later. Fore a more complete discussion, we address the
reader to the monograph \cite{tink}.

\appendix
\section{Magnetic properties of type I superconductors}

As  is well known, superconductors tend to expel magnetic fields
from their interior ({\it Meissner effect}). The reversibility of
this effect implies the existence of a well defined critical
magnetic field $H_c$ that destroys superconductivity, via a
first-order phase transition. If we denote by $e(T)$     the
difference between the  zero-field  Helmoltz free energies (per
unit volume) $f_n(T)$ and $f_s(T)$  in the normal (n) and the
superconducting (s) states respectively, i.e. \be e(T)
:=f_n(T)-f_s(T)\;, \ee ($e(T)$ is called the condensation energy
of the superconductor) it can be shown that the value of $H_c$ is
related to $e(T)$ by the relation \be \frac{H^2_c(T)}{8
\pi}=e(T)\;.\label{hterm}\ee Upon using the well known empirical
parabolic law for $H_c$ : \be H_c(T) \approx
H_c(0)\,(1-t^2),\;\label{parab}\ee where $t=T/T_c$ with $T_c$  the
critical temperature,  it follows  from Eq. (\ref{hterm}) that the
condensation energy has the following dependence on $t$: \be
e(T)=e(0)\,(1-t^2)^2\;,\label{econtd}\ee where \be
e(0)=\frac{H_c^2(0)}{8 \pi}\;.\label{ezero} \ee Another empirical
law of interest \cite{lewis} is the one giving the approximate
dependence of $H_c(0)$ on $T_c$ (valid for type I superconductors)
\be H_c(0) \propto T_c^{1.3}\;,\label{emptc}\ee and then by using
Eq. (\ref{ezero}) we obtain an estimate of the dependence of
$e(0)$ on the critical temperature of the superconductor in the
form \be e(0) \propto T_c^{2.6}\;.\label{ectc}\ee

In the case of a thin film, of thickness $D \ll \lambda,\xi$ (with
$\lambda$ the penetration depth and $\xi$ the correlation length),
placed in a parallel magnetic field, expulsion of the magnetic
field is incomplete, and consequently the critical field increases
from $H_c$ to $H_{c \|}$.  Unlike the bulk case, the transition to
the normal phase is of second order,  since sufficiently thin
films and use of the Ginzburg--Landau theory show that $H_{c \|}$
is proportional to $H_c(T)$ \be H_{c \|}(T) =
\rho(T)\,H_c(T)\;.\label{appe}\ee The coefficient $\rho$ has the
approximate expression \be \rho \approx\sqrt{24}\;
\frac{\lambda}{D}\left(1+\frac{9 D^2}{\pi^6
\xi^2}\right)\;,\label{hfilm}\ee where the second term inside the
brackets accounts for surface nucleation. Upon  using Eq.
(\ref{appe}) to express $H_c$ in terms of $H_{c \|}$, we can
recast Eq. (\ref{hterm})   in the form \be \frac{1}{8
\pi}\,\left(\frac{H_{c \|}(T)}{\rho}\right)^2 \,=
e(T)\;.\label{hfilm}\ee Recalling that the temperature dependence
of $\lambda$ can be approximately described by the law \be
\lambda(T) \approx \lambda(0)[1-(T/T_c)^4]^{-1/2}\;,\ee it follows
from Eq. (\ref{hfilm}) that, close to $T_c$, $H_{c \|}$ approaches
zero like $\sqrt{1-T/T_c}$.

In the case of ultrathin films (with thicknesses of a few nm) the
magnetic behavior can be quite different because the orbital
currents are suppressed, and the transition to the normal state is
driven by coupling of the magnetic field to the electronic spins.
Under such circumstances, the critical field transition can be
strongly hysteretic, below a tricritical point \cite{weh}.
However, close to $T_c$, there is no hysteresis, and if the films
are very pure it is still possible to observe very sharp
transitions, with well defined critical fields (defined to within
a few parts in a thousand) \cite{adam2}.

It is  important to point out that the critical temperature $T_c$
of a thin film does not coincide in general with  the bulk value,
and  it depends on the film thickness and on the preparation
procedure. For example, in the case of thin Beryllium films one
can have a $T_c$ as high as $9$ K, to be contrasted with the bulk
value of $24$ mK \cite{takei}. In such cases, the condensation
energy of the film should not be expected to be the same as for
the bulk material. However, one can estimate it by using the BCS
formula for the  condensation energy (per unit volume) at absolute
zero: \be  e(0)=\frac{1}{2}\, N(0)\, \Delta^2(0)\label{park}\;,\ee
where $N(0)$ is the density of electronic states of one spin in
the normal metal at the Fermi surface \cite{parks}. If we assume
that the gap $\Delta(T)$ is always given by the approximate BCS
formula \cite{carless} \be 2\Delta(T)=3.528 \,k \, T_c\, \sqrt{1-
\frac{T}{T_c}}\,\left(0.9963+0.7735 \,
\frac{T}{T_c}\right)\;,\label{gap}\ee    and that $N(0)$ for the
film is not much different from the bulk value, we obtain from Eq.
(\ref{park}) \be \frac{ e(0)|_{\rm film}}{ e(0)|_{\rm
bulk}}=\frac{T_c^2|_{\rm film}}{T_c^2|_{\rm bulk}}.\ee If we
further assume that the temperature dependence of the condensation
energy of a film is still of the   form Eq. (\ref{econtd}),  the
same ratio is found  also for the free energies  at finite values
of $t$.

\section{High-frequency electrodynamics}

As  is well known, superconductors show finite dissipation when
traversed by alternating currents and/or time-varying
electromagnetic fields. At   frequencies $\omega$ much smaller
than the energy gap, $\hbar \omega \ll \Delta(T)$, a qualitative
description of the superconductor response is provided by the
simple Casimir--Gorter two-fluid model, which leads to the
following expression for the real part $\sigma'(\omega)$ of the
complex conductivity $\sigma(\omega)$: \be \sigma'(\omega)=(\pi
n_s e^2/2m)\; \delta(\omega)+(n_n e^2 \tau_n/m) \; (1+\omega^2
\tau_n^2)^{-1}\;, \label{casgor}\ee where $\delta(\omega)$ is the
Dirac delta function, $m$ and $e$ denote the electron mass and
charge, $n_s$ and $n_n$ are the temperature dependent densities of
the superconducting and normal electron respectively, and $\tau_n$
is the residual relaxation time for the normal electrons, as
determined by the impurities present in the sample. As we see, the
normal fluid contribution implies a {\it nonzero} dissipation {\it
at all nonzero frequencies}. The delta function contribution,
proportional to $n_s$ is a dc contribution, ensuring fulfillment
of the oscillator strength sum rule: \be \int_0^{\infty} d \omega
\;\sigma'(\omega)= \frac{\pi n e^2}{2m}\;,\label{srule}\ee where
$n=n_s+n_n$ is the total electron density.

An accurate description of the complex conductivity of a BCS
superconductor, for photon energies up to and larger than (twice)
the gap, requires that one uses the full   Mattis--Bardeen theory
of electric conduction, which is valid also in the domain of the
anomalous skin effect \cite{mattis}.  Taking full account of space
dispersion, this theory gives rise to an expression of the complex
conductivity $\sigma_s(\omega, \bf{q})$ that depends also on the
spatial wave vector $\bf{q}$. In the case of a BCS conductor at a
temperature $T<T_c$, and in the local limit ${\bf q} \rightarrow
0$,  the resulting expression of $\sigma'_s(\omega)$ can be
written in a form analogous to Eq. (\ref{casgor}): \be
\sigma'_s(\omega)=\kappa \,\delta(\omega)+\hat{\sigma}'_s(\omega)
\;.\label{bcssum}\ee  For $\omega > 0$, $\hat{\sigma}'_s(\omega)$
reads as \cite{berl} \be \hat{\sigma}'_s(\omega)=\frac{\hbar \,
n\, e^2}{2 m \omega
\tau_n}\left[\int_{\Delta}^{\infty}dE\,J_T+\theta(\hbar \omega-2
\Delta)\int_{\Delta -\hbar \omega}^{-\Delta}dE\,\,J_D
\right]\;,\label{bcscon}\ee where
\begin{eqnarray}
  J_T &:=& \,g(\omega,\tau_n,E)\,\left[\tanh \frac{E+\hbar \omega}{2 k T}-
  \tanh \frac{E}{2kT}\right] \\
  J_D &:=& -g(\omega,\tau_n,E)\;\tanh \left(\frac{E}{2k
  T}\right)\;,
\end{eqnarray}
with $k$ the Boltzmann constant. Defining \be P_1:=\sqrt{(E+\hbar
\omega)^2-\Delta^2}\;,\;\;\;\; P_2:=\sqrt{E^2-\Delta^2}\;, \ee the
function $g(\omega,\tau_n,E)$ is
\begin{eqnarray}
  g &:=& \left[1+\frac{E(E+\hbar \omega)+\Delta^2}{P_1 P_2}\right]\frac{1}{(P_1-P_2)^2+(\hbar/\tau_n)^2}
  \nonumber\\
   &-&\left[1-\frac{E(E+\hbar \omega)+\Delta^2}{P_1 P_2} \right]\frac{1}{(P_1-P_2)^2+(\hbar/\tau_n)^2}
  \;.\nonumber
\end{eqnarray}
The  coefficient $\kappa$ of the delta function  in Eq.
(\ref{bcssum}) is again determined so as to satisfy the sum rule
Eq. (\ref{srule}) and can be computed exactly according to
\cite{berl}
\begin{widetext}\be  \kappa=\frac{\pi n e^2}{m}
\;  \left[\frac{\pi \tau_n \Delta}{\hbar} \tanh \frac{\Delta}{2 k
T}-4 \Delta^2 \int_{\Delta}^{\infty} dE\;\frac{\tanh(E/2 k
T)}{\sqrt{E^2-\Delta^2}[4(E^2-\Delta^2)+(\hbar/\tau_n)^2]}\right]\;.\label{delta}\ee
\end{widetext}
The conductivity $\sigma'_s(\omega)$ in Eq. (\ref{bcssum}) can be
thought of as the sum of three contributions: a $\delta$ function
at the origin, a broad thermal component that diverges
logarithmically at $\omega=0$ and a direct absorption component,
with an onset at $2 \Delta(T)$ (see the second in Ref.
\cite{berl}). At any $T<T_c$, complete specification of
$\sigma'_s(\omega)$ requires three parameters: besides the free
electron density $n$ (or equivalently the square of the plasma
frequency $\Omega_n^2=4 \pi n e^2/m$) that provides the overall
scale of $\sigma'_s$, and the relaxation time for the normal
electrons $\tau_n$, both of which already occur in the simple
Drude formula Eq. (\ref{drude}) below, $\sigma'_s(\omega)$ only
depends on one extra parameter, i.e. the gap $\Delta$. In fact,
after the frequencies are expressed in reduced units $x=\hbar
\omega/(2 \Delta)$, the expression inside the brackets on the
r.h.s. of Eq. (\ref{bcscon}) becomes a function solely of $x,\;
2\Delta/(k\,T)$ and the so-called impurity parameter
$y=\hbar/(2\Delta \tau_n)$. We point out that this expression for
$\sigma'_s$ is valid for arbitrary relaxation times $\tau_n$, i.e.
for arbitrary mean free paths, and in particular it holds in the
so-called impure limit $y=2\Delta/(\hbar \tau_n) \gg 1$, where the
effects of non-locality become negligible.
\begin{figure}
\includegraphics{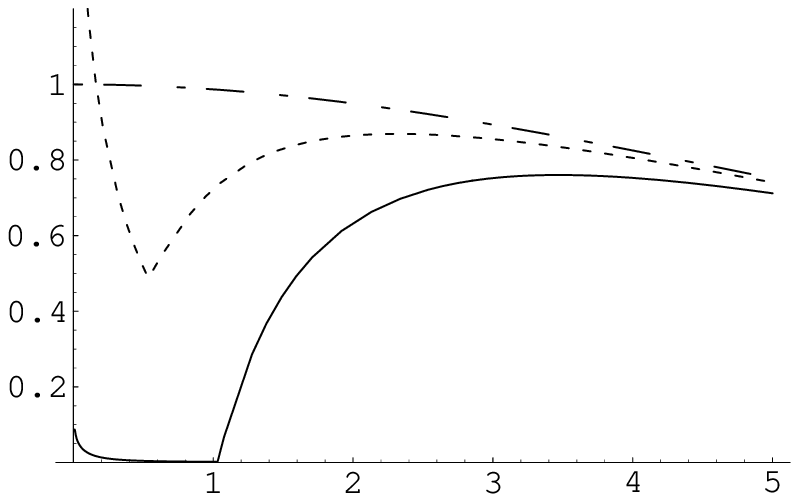}
\caption{\label{plotssigma} Plots of $   m \,\sigma'_s(\omega)/( n
e^2 \tau_n)$, for $T/T_c=0.3$ (solid line), $T/T_c=0.9$ (dashed
line) and $T=T_c$ (point-dashed line). On the abscissa, the
frequency $\omega$ is in   reduced units $x_0=\hbar \omega/(2
\Delta(0))$, and $y_0=2 \Delta(0)/\tau_n \simeq 8.7 $.}
\end{figure}
We recall that the gap $\Delta$ is a temperature dependent
quantity.  In our numerical computations, we used for it the
approximate formula Eq. (\ref{gap}).

We  point out that at fixed $\omega$ for $T \rightarrow T_c$, as
well as at fixed $T<T_c$ for $x\rightarrow \infty$,
$\sigma'_s(\omega)$ approaches the Drude expression
$\sigma_D'(\omega)$ \be \sigma'_D(\omega)=\frac{1}{4 \pi}
\frac{\Omega^2\; \tau}{1+\omega^2 \tau^2}. \label{drude}\ee  The
convergence of $\sigma'_s(\omega)$ to $\sigma_D'(\omega)$ in the
frequency domain is in fact very fast, and already for $x$ of
order 10 or so $\sigma'_s$ becomes undistinguishable from
$\sigma_D'$, in accordance with experimental findings
\cite{glover}. In Fig. 5, we show the plots of $ \sigma'_s(\omega)
\,m/ (n e^2 \tau_n)$, for $T/T_c=0.3$, $T/T_c=0.9$ and $T=T_c$.
The curves are computed for $y_0=2 \Delta(0)/\tau_n \simeq 8.7 $.
Frequencies are measured in reduced units $x_0=\hbar \omega/(2
\Delta(0))$.

As we see from  Eqs. (\ref{denren}) and (\ref{fint}) the
expression for the variation of Casimir free energy involves the
permittivity of the superconducting film $\epsilon_s(i \zeta)$ for
complex frequencies $i \zeta$.  Now, $\epsilon_s(i \zeta)$ cannot
be given in closed form, and we  have estimated  it as follows.
First, by using well known dispersion relations, we express
$\epsilon_s(i \zeta)$ in terms of the imaginary part
$\epsilon''_s(\omega)$ of the dielectric permittivity
$\epsilon_s(\omega)=\epsilon'_s(\omega)+i\, \epsilon''_s(\omega)$
at real frequencies, i.e. \be \epsilon_s(i \zeta)-1=\frac{2}{\pi}
\int_0^{\infty} d\omega \frac{\omega
\,\epsilon''_s(\omega)}{\zeta^2+\omega^2}\;.\label{disp}\ee Upon
using now the  standard relation connecting the imaginariy part of
the electric permittivity to the real part of the complex
conductivity $\sigma$ of a metal: \be \epsilon''(\omega)= \frac{4
\pi}{\omega}\,\sigma'(\omega)\;,\label{cond} \ee we can rewrite
the r.h.s. of  Eq. (\ref{disp})  as \be \epsilon_s(i \zeta) =1+8
\int_{0}^{\infty} d \omega
\,\frac{{\sigma}'_s(\omega)}{\zeta^2+\omega^2}\;.\label{dispmet}\ee
We now take for $\sigma'_s(\omega)$ the semianalytical BCS formula
given in Eqs. (\ref{bcssum}-\ref{delta}). Upon plugging Eq.
(\ref{bcssum}) into Eq. (\ref{dispmet}) we then obtain the
following formula for $\epsilon_s(i \zeta)$: \be \epsilon_s(i
\zeta) =1+8 \int_{0}^{\infty} d \omega
\,\frac{\hat{\sigma}'_s(\omega)}{\zeta^2+\omega^2}\;+\; \frac{4
\kappa}{\zeta^2}\;, \label{dispsup}\ee where $\kappa$ is given in
Eq. (\ref{delta}).  For the purpose of   a numerical evaluation of
Eq. (\ref{dispsup}), it is however convenient to rewrite it in a
different form. We recall that the coefficient $\kappa$ is defined
so as to satisfy the sum rule Eq. (\ref{srule}). Now, upon
substituting the expression  for $\sigma'_s(\omega)$ Eq.
(\ref{bcssum}) into Eq. (\ref{srule}) we obtain \be
\frac{\kappa}{2}+ \int_0^{\infty} d \omega
\;\hat{\sigma}'_s(\omega)= \frac{\pi n e^2}{2m}\;.\ee However, the
sum rule holds also in the normal state of the film, and thus we
have \be \int_0^{\infty} d \omega \;{\sigma}'_D(\omega)= \frac{\pi
n e^2}{2m}\;,\ee where $\sigma'_D(\omega)$ (see Eq. (\ref{drude}))
is the expression derived from Eq. (\ref{drper}), by using Eq.
(\ref{cond}). Upon equating the l.h.s. of the above two Equations,
we arrive at the following expression for $\kappa/2$: \be
\frac{\kappa}{2}=\int_0^{\infty} d \omega
\;({\sigma}'_D(\omega)-\hat{\sigma}'_s(\omega))\;.\ee Upon
plugging this expression for $\kappa$ into Eq. (\ref{dispsup}),
and by using the dispersion relation for $\sigma'_D$, i.e. \be
\epsilon_D(i \zeta)-1=   8 \int_{0}^{\infty} d \omega \,\frac{
{\sigma}'_D(\omega)}{\zeta^2+\omega^2}\;,\ee   it is easy to
obtain the  final equation \be \epsilon_s(i \zeta) = \epsilon_D(i
\zeta)- \frac{8}{\zeta^2} \int_{0}^{\infty} d \omega
\,\frac{\omega^2(\hat{\sigma}'_s(\omega)-\sigma'_D(\omega))}{\zeta^2+\omega^2}\,,\label{diffeps}\ee
that we have used in our numerical computations. It has the virtue
of expressing the variation in the film permittivity $\epsilon_s(i
\zeta)- \epsilon_D(i \zeta)$ across the transition, as an integral
involving the  variation of conductivity
$\hat{\sigma}'_s(\omega)-\sigma'_D(\omega)$.  Indeed,
$\hat{\sigma}'_s(\omega)-\sigma'_D(\omega)$  is appreciably
different from zero only for frequencies $\omega$ of   order of a
few times $ k\,T_c/\hbar$, and goes to zero like $1/\omega^{3}$
for large $\omega$'s, and this ensures rapid convergence of the
integral on the r.h.s. of Eq. (\ref{diffeps}). By looking at the
plots in Fig. (\ref{plotssigma}), we see that, for frequencies of
order $2 \Delta/\hbar$ (apart from a very narrow region near the
origin, where $\hat{\sigma}_s(\omega)$ has a logarithmic
behavior), $\hat{\sigma}_s(\omega) < \hat{\sigma}_n(\omega)$, and
therefore it follows from Eq. (\ref{diffeps}) that $\epsilon_s(i
\zeta)> \epsilon_n(i \zeta)$. This relation is fully confirmed by
our numerical computations.

\end{document}